\documentclass[conference]{IEEEtran}
\usepackage{graphicx}
\usepackage{float}
\usepackage{multirow}
\usepackage{subcaption}
\ifCLASSOPTIONcompsoc
  \usepackage[nocompress]{cite}
\else
  \usepackage{cite}
\fi

\ifCLASSINFOpdf
 
\else
 
\fi

\usepackage{color}
\usepackage[table,xcdraw]{xcolor}
\usepackage{graphicx}
\usepackage{amssymb}
\usepackage{amsthm}
\usepackage{amsmath}
\usepackage{algorithm}
\usepackage{algorithmic}

\usepackage{hyperref}
\hypersetup{
	bookmarks=false,
	colorlinks=false,       
	linkcolor=blue,          
	citecolor=green,        
	filecolor=magenta,      
	urlcolor=cyan         
}

\usepackage{lipsum}
\usepackage{multicol}
\usepackage{graphicx}
\usepackage[acronym]{glossaries}
\newacronym{gmft}{GMFT}{Google Mobile-Friendly Test Tool}
\newacronym{psit}{PSIT}{Google PageSpeed Insights Tool}
\newacronym{xbis}{XBIs}{Cross Browser Issues}
\newacronym{pdg}{PDG}{Property Dependence Graph}
\newacronym{sm}{SM}{Segment Model}
\newacronym{ism}{ISM}{Intra-Segment Model}
\newacronym{dom}{DOM}{Document Object Model}
\newacronym{pso}{PSO}{Particle Swarm Optimization}

\makeatletter
\newcommand{\newlineauthors}{%
  \end{@IEEEauthorhalign}\hfill\mbox{}\par
  \mbox{}\hfill\begin{@IEEEauthorhalign}
}
\makeatother

\newcounter{todocounter}

\usepackage{amsmath}
\usepackage{amssymb}

\title{A title}
\author{\IEEEauthorblockN{Thanh Le-Cong}
\IEEEauthorblockA{School of Information and\\Communication Technology\\
Hanoi University of Science and Technology\\
Hanoi, Vietnam\\
thanh.ld164834@sis.hust.edu.vn}
\and
\IEEEauthorblockN{Xuan Bach D. Le}
\IEEEauthorblockA{School of Computing and \\ Information Systems\\
University of Melbourne \\
Melbourne, Australia\\
bach.le@unimelb.edu.au}
\and
\IEEEauthorblockN{Quyet Thang Huynh \IEEEauthorrefmark{2} and Phi Le Nguyen}
\IEEEauthorblockA{School of Information and\\Communication Technology\\
Hanoi University of Science and Technology\\
Hanoi, Vietnam\\
\{thanghq, lenp \}@soict.hust.edu.vn}}

\begin{document}

\title{Usability and Aesthetics: Better Together for Automated Repair of Web Pages}

\maketitle
\begingroup\renewcommand\thefootnote{\IEEEauthorrefmark{2}}
\footnotetext{Corresponding author}
\begin{abstract}
With the recent explosive growth of mobile devices such as smartphones or tablets, guaranteeing consistent web appearance across all environments has become a significant problem.  This happens simply because it is hard to keep track of the web appearance on different sizes and types of devices that render the web pages. Therefore, fixing the inconsistent appearance of web pages can be difficult, and the cost incurred can be huge, e.g., poor user experience and financial loss due to it. Recently, automated web repair techniques have been proposed to automatically resolve inconsistent web page appearance, focusing on improving usability. However, generated patches tend to disrupt the webpage's layout, rendering the repaired webpage aesthetically unpleasing, e.g., distorted images or misalignment of components.

In this paper, we propose an automated repair approach for web pages based on meta-heuristic algorithms that can assure both usability and aesthetics. The key novelty that empowers our approach is a novel fitness function that allows us to optimistically evolve buggy web pages to find the best solution that optimizes both usability and aesthetics at the same time. Empirical evaluations show that our approach is able to successfully resolve mobile-friendly problems in 94\% of the evaluation subjects, significantly outperforming state-of-the-art baseline techniques in terms of both usability and aesthetics.  


\end{abstract}


\IEEEpeerreviewmaketitle

\section{Introduction}
\label{sec:introduction}
In our modern world, the World Wide Web (WWW) has become one of the most popular sources
of information  \cite{eMarketer, Consumer}. This information is often accessible through web pages via numerous types of devices such as smartphones and tablets and via various web browsers such as
Chrome, Safari, Firefox, Internet Explorer, and many more.
The variety of devices and browsers, on the one hand, allows various ways to access information, but on the other hand, poses several profound challenges,
among which guaranteeing consistent web appearance across all environments is one of the most important problems.

\par  In practice, developers often have to ensure that web pages' appearance is amenable to various environments. Specifically, there are two popular types of web appearance issues that developers usually concern with, including usability and aesthetics of web pages across sizes and types of different devices. Usability issues include bugs related to font size, which affect readability, the margin of touchable elements, the presence of navigation or content that overflows the device's viewport, etc. Aesthetic issues concern relative proportions and positioning of elements on the page. To ensure the absence of these issues, developers often have to test web pages and fix any discovered errors manually. This common practice, however, is very time-consuming and very expensive due to the enormous amount of environments that need to be considered. For example, these environments may refer to multiple browsers such as Chrome, Firefox, or multiple devices such as smartphones, tablets. Automated techniques that can help developers cope with automatic testing and repairing bugs on web pages are thus of tremendous value.

Recent years have seen a pragmatic progress on automated web repair techniques proposed to resolve usability and aesthetic issues on web pages ~\cite{mfix, Mahajan:2017:ARL:3092703.3092726}. 
These approaches generally follow three stages: bugs discovery, localization, and repair. The first stage, i.e., bugs discovery, can be performed via readily available tools such as \gls{gmft} \cite{gmft}, \gls{psit} \cite{PSIT} or Bing \cite{Bing}. The latter two stages, i.e., localization and repair, are often more challenging with less tool support. Techniques like MFix~\cite{mfix} achieve this often via a genetic programming approach that gradually
generates repair candidates that improve usability over time. Although proposed repair algorithms can help to improve web pages' appearance, since they mainly focus on usability while
letting aesthetics be best-effort, they may not provide a patching solution that can guarantee good usability and aesthetics simultaneously. 

Machine-patched pages may suffer from a poor aesthetic and result in an undesirable layout. For example, Figure \ref{fig:mfix_repair} depicts a repaired
page by state-of-the-art MFix~\cite{mfix} across four different runs. From the figure, we can clearly see the instability and imbalance of
MFix. The first two patches have layout distortion as some symbols and text are overlapped. Compared to the original page, the third patch does not change, meaning it does not manipulate the original page to reflect developers' intent. The fourth patch is even more severely problematic as it has cluttered navigation. 

\begin{figure*}
\centering
\begin{subfigure}[H]{0.19\textwidth}
\includegraphics[width=\textwidth]{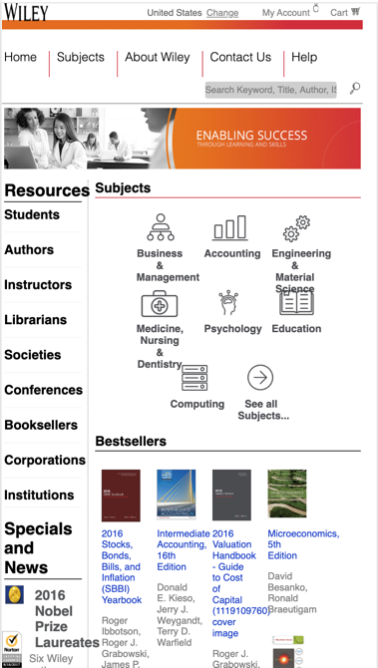}
\caption{Run 1}
\end{subfigure}
\begin{subfigure}[H]{0.19\textwidth}
\includegraphics[width=\textwidth]{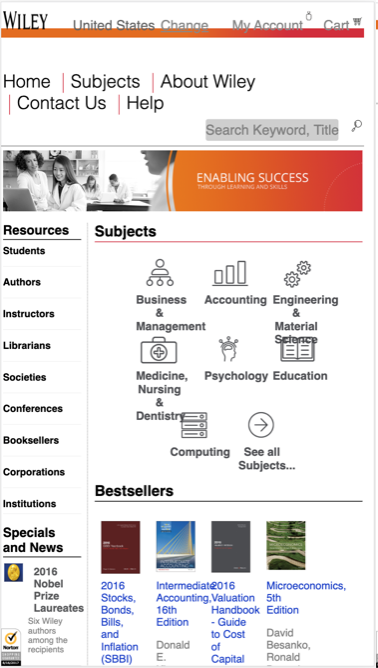}
\caption{Run 2}
\end{subfigure}
\begin{subfigure}[H]{0.19\textwidth}
\includegraphics[width=\textwidth]{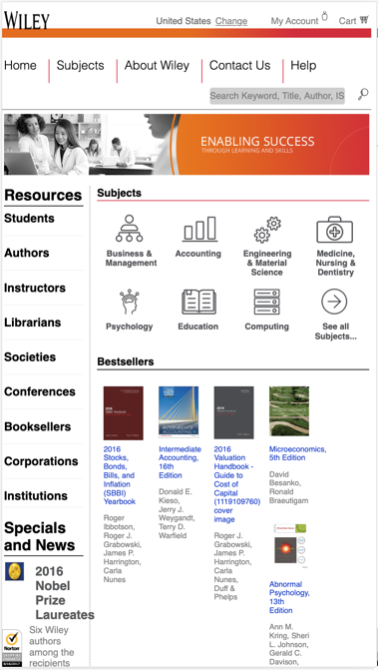}
\caption{Run 3}
\end{subfigure}
\begin{subfigure}[H]{0.38\textwidth}
\includegraphics[width=\textwidth]{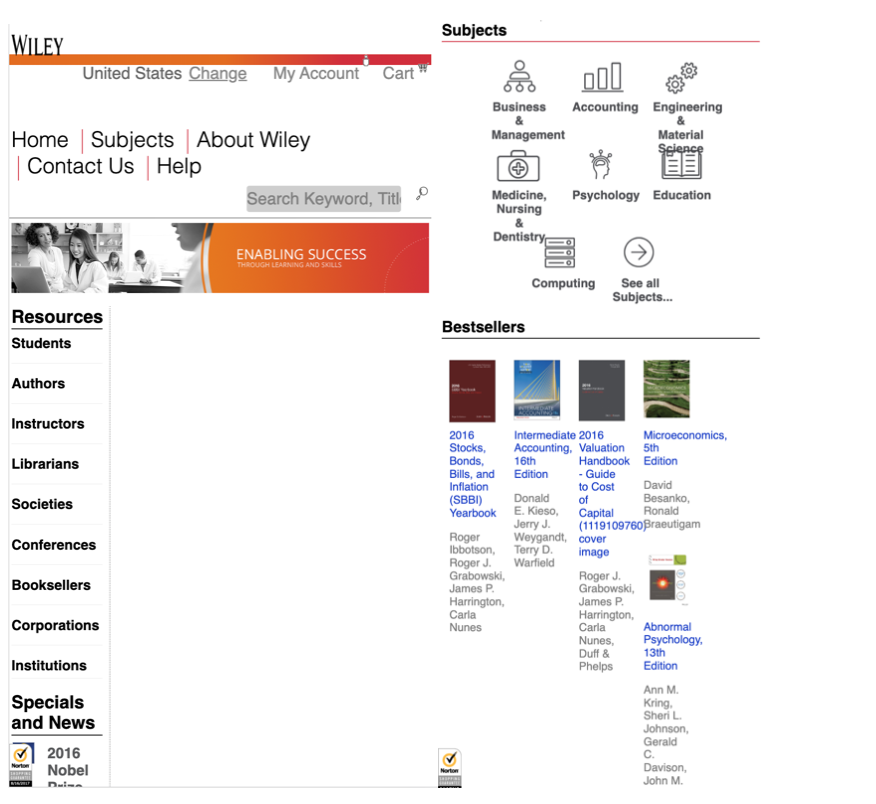}
\caption{Run 4}
\end{subfigure}
\caption{Repaired page across several runs by MFix \label{fig:mfix_repair}}
\end{figure*}
\par In this work, we focus on automated repair of web pages, considering both usability and aesthetics. 
To assess usability, we use PSIT \cite{PSIT}, which rates how friendly a web page is based on a scale from 0 to 100; the higher the score, the more usable the web page is. To assess aesthetics, we follow a common metric proposed in~\cite{mfix}, which calculates the alignment between components in a web page. Based on these metrics, we propose a new fitness function that optimizes both the usability and aesthetics scores simultaneously. We propose two repair algorithms to search for a solution that not only addresses mobile friendliness problems but also ensures a pleasing layout.
In the first algorithm, we focus on the accuracy of a solution with a high level of usability and aesthetics at a reasonable time cost. In the second algorithm, we compensate the accuracy for the running time. Specifically, we try to reduce the running time as much as possible while maintaining an acceptable level of usability and aesthetics. The first algorithm follows \gls{pso} \cite{kennedy95particle}, and the second is based on Tabu Search ~\cite{glover1989tabu}. We compare our approach against MFix~\cite{mfix} -- the state-of-the-art of mobile-friendly web repair. Similar to MFix, our algorithms generate and improve repair candidates over several generations. Different from MFix, at the core of our algorithms, we design a fitness function that optimizes for both usability and aesthetics.

\par We evaluated our approaches and MFix on 38 web pages, which is provided by Mahajan et al. \cite{mfix} on two tasks: mobile-friendliness (usability) and aesthetic layout (aesthetics). 
Experiment results show that our approaches significantly outperform MFix. Notably, our \gls{pso}-based repair algorithm outperforms MFix on 29 of 38 web pages in the aesthetic layout task and on 26 of 38 web pages in the mobile-friendliness task. Also, our Tabu search-based algorithm shows superior efficiency despite being less effective than the \gls{pso}-based algorithm that we proposed. Finally, we also evaluate different parameter settings for \gls{pso} and Tabu Search to show their impact. Experiments with the Wilcoxon signed-rank test~\cite{Wilcoxon1992} show that the superiority of our approaches over MFix is significant.
\par In summary, our contributions include:
\begin{itemize}
        \item We propose a new fitness function that optimizes both the usability and aesthetics scores to guarantee generated solutions that not only address mobile friendliness problems but also ensure a pleasing layout. 
        \item We proposed two algorithms, namely PBRA and TERA, to search for solutions based on the proposed fitness function. The first algorithm focuses on the accuracy of solutions, while the second algorithm reduces the running time. 
        \item We demonstrate that our approach significantly outperforms the current state-of-the-art MFix on aesthetic layout and mobile-friendliness tasks. 
\end{itemize}
\par The remainder of the paper is organized as follows.
Section \ref{sec:background} introduces the overview of an automated web repair process. We present the mathematical formulation of the problem in Section \ref{sec:formulation} and describe our proposed algorithms in Section \ref{sec:proposal}. Section \ref{sec:evaluation} presents the experimental results, and Section \ref{sec:conclusion} concludes the paper.

\section{Related Work} \label{sec:relatedwork}
Recently, there has been a growing interest in automated fixing of presentation issues in web pages. Mahajan et al. propose XFix ~\cite{Mahajan:2017:ARL:3092703.3092726} technique that repairs presentation failures
arising from the inconsistencies in the rendering of a website across
different browsers, i.e., layout  Cross Browser Issues (XBIs). XFix assumes that one of the browsers presents the correct presentation of the page, which
must be rendered the same in other browsers. In this work, we focus on the usability, and aesthetic of web pages, which is different from the XBIs addressed by XFix. That is, there is no correct reference available to the repair process in our approach. Mahajan et al. propose MFix ~\cite{mfix} technique which is a novel automated approach for repairing mobile-friendly problems in web pages. They also provided effective metrics to assess the aesthetic usability of web pages. However,  MFix mainly focuses on usability while letting the aesthetic best effort. Mahajan et al. recently proposed $\text { IFix }^{++}$ \cite{mahajan2021effective},  a search-based technique to automatically repair Internationalization Presentation Failures in web application. However, this work only fixes layout disruption, while our work focuses on both usability and layout disruption at the same time.

Cassius et al.~\cite{panchekha2016automated}, proposes to use automated
reasoning for debugging and repairing buggy CSS. The
technique takes as input faulty source lines in CSS files
and user-provided examples that can be used to guide the
repair synthesis. These inputs are, however, not readily available in the problem domain that our work addresses. PhpRepair~\cite{samimi2012automated} and PhpSync~\cite{nguyen2011auto} detect and repair HTML syntax issues. Wang et
al.~\cite{wang2012automating} propose to use static and dynamic analysis together
for repairing web applications. This is achieved by propagating a generated presentation fix to the server-side source code. 

Less relevant to our work in this paper is a recent research effort in automated repair of source code in Java or C programming languages. Le Goues et al. spark the exciting pioneering work on automated fixing of bugs in C programs~\cite{le2011genprog}. The research in automated bug fixing has since then seen a plethora of proposed techniques for fixing bugs in both C and Java programs, using an array of techniques including search-based software engineering and formal methods, etc~\cite{le2016history, nguyen2013semfix, mechtaev2016angelix, le2017s3}. These approaches assume the existence of a test suite to guard against the correctness of the programs under repair. Similar to the search-based program repair, e.g.,~\cite{le2011genprog, le2016history}, our work uses evolutionary search to find the best candidates. However, different from them, we do not rely on a test suite for patch validation, but instead, we use readily available tools that can automatically assess usability and aesthetics to evaluate repair candidates. 

Overall, despite the recent interest in repairing web pages, there are still more works needed to push the boundary of this research area further. Our work attempts to enhance the existing state-of-the-art work on web page repair. The focus of our work is on repairing web pages while preserving usability and aesthetics altogether. By doing this, our work is quite unique on its own to tackle the aforementioned problem. In the next section, we describe the details of our proposed framework. 
\section{Our Framework}\label{sec:background}

Our repair framework consists of three phases: segmentation, localization, and repair. The main novelty of our framework lies in the third phase, i.e., repair, while the first two phases are inspired by~\cite{mfix}. Below, we describe each phase in more detail.

The segmentation phase groups web pages into regions.
Each region is a set of HTML elements that should be repaired together to guarantee the visual consistency of the repaired page. 
This can be achieved by readily available tools, such as VIPS \cite{vips}, Block-o-Matic \cite{bom}, correlation clustering \cite{cluster}, and cluster-based partition \cite{repair}. In this work, we leverage an automated clustering-based partitioning algorithm proposed by Romero et al. \cite{repair} as our segmentation tool. This approach is based on the  Document Object Model (DOM) tree of the page, in which each leaf element of the DOM tree is assigned to its segment. Then, the segments are merged if they are close enough, i.e., their distance measured by the average depth of leaves in the DOM tree is below a predefined threshold. 

The localization phase consists of two steps. 
The first step is for classifying issues associated with problematic segments that should be repaired.
There are five main types of issues, including font-sizing, tap target spacing, content sizing, viewport configuration, and flash usage as reported by popular mobile testing tools \cite{PSIT}, \cite{Bing}. In this paper, we only focus on addressing the first three of these problems following MFix \cite{mfix}. The result of this step is a set of tuples with the form of $\left \langle s,T \right \rangle$, where $s$ is a potentially problematic segment and $T$ is the set of issues associated with $s$. 
Note that, $T$ is a subset of \emph{\{font-sizing, tap target spacing, content sizing\}}.
To achieve this, Google Mobile-Friendly Test Tool (GMFT) can be used.
Next, the second step determines CSS properties that may need to be adjusted for patching the issues of the problematic segments.
For example, a font-sizing issue may be patched by adjusting \emph{font-size, line-height, width, height} properties. As all the elements in the same segment should have a reciprocal relationship, their CSS properties should be adjusted together. 
To address this problem, MFix~\cite{mfix} proposed a structure named \gls{pdg}, which represents the relationship between elements in a segment with respect to a given issue type. 
Specifically, each node of a \gls{pdg} graph corresponds to an HTML element in the segment.
Two nodes in the \gls{pdg} are connected if there exists a dependency relationship between them. There is a function $M$ that maps each edge of the \gls{pdg} to the ratio of between the values of the CSS properties associating with the two end nodes of the edge. 
Intuitively, if we know the value corresponding to the root node of a \gls{pdg}, we can deduce the value corresponding to the other nodes by using function $M$.
This results in a set of segments and their corresponding \gls{pdg}s. 

The last phase, i.e., repair, is to determine the optimal values of CSS properties for repairing all the found issues. 
In this paper, we focus on this phase and propose two algorithms which will be described in the following sections.

\section{Formulation and Fitness Function}
\label{sec:formulation}
The core of our algorithms is how we define a fitness function to search for the best solution during the evolution of repair candidates. A well-designed fitness function can lead to better repair candidates, and thus, we carefully crafted the fitness function as explained below. We first give the detailed formula for the fitness function that we designed and then explain the high-level idea following the design of the function.

Given the potentially problematic segments, their issues, and the corresponding \gls{pdg} identified by the segmentation and localization algorithms described in the previous section, our objective is to determine the optimal patching solution to address all the issues. 
Let $\mathbb{S}$ denote the set of all segments, $\mathbb{S} = \{S_1, ..., S_n\}$.
For each $S_i$ ($i = \overline{1,n}$), let $I_i$ be the set of issues associated with $S_i$.
Let $I_{i}^{j}$ be an issue belonging to $I_i$, then $j \in \{1, 2, 3\}$, where $I_{i}^{1}, I_{i}^{2}, I_{i}^{3}$ correspond to issues related to font-size, tap target spacing and content sizing, respectively. 
Let $\operatorname{PDG}{\left \langle S_i,I_{i}^{j} \right \rangle}$ denote the \gls{pdg} corresponding to $S_i$ and $I_{i}^{j}$. 
As mentioned in Section \ref{sec:background}, to fix issue $I_{i}^{j}$, we just need to determine the value of CSS properties associated with the root node of $\operatorname{PDG}{\left \langle S_i,I_{i}^{j}\right \rangle}$ (then, the other elements can be adjusted by using \gls{pdg}). 
Our problem can be mathematically formulated as follows.  

\noindent \textbf{Input}
\begin{equation*}
S_i, I_{i}^{j}, \operatorname{PDG}\left \langle S_i,I_{i}^{j} \right \rangle ~~~(i=\overline{1,n}; j\in\{1, 2, 3\})
\end{equation*}

\noindent \textbf{Output} 
\begin{equation*}
\mathbb{X}= \left \{ X_{i}^{j} \right \} ~~~(i=\overline{1,n}; j\in{1, 2, 3}),
\end{equation*}
where $X_i^{j}$ denotes the value of CSS properties associated with the root node of ${\left \langle S_i,I_{i}^{j} \right \rangle}$. \\

\noindent \textbf{Objectives}
\begin{equation*}
\begin{matrix}
\text{Maximize} & \text{Usability score~} U(\mathbb{X}) \\
\text{Minimize} & \text{Aesthetic score~} A(\mathbb{X})
\end{matrix}
\end{equation*}
where $U(\mathbb{X})$ indicates the mobile friendliness of the page which can be measured by using \gls{psit};  the aesthetic score quantifies the layout differences between the original page and a transformed page to which a candidate patch has been applied. The aesthetic score is measured based on \gls{sm} and \gls{ism}, where:
\begin{itemize}
    \item A Segment Model (SM) is defined as a directed complete graph where the nodes are the segments in the segmentation phase and the edges are labeled by one of four relationships: (1) intersection, (2) containment, (3) directional (i.e., above, below, left, and right). 
    \item A Intra-segment Model (ISM) is similar to SM but is built for each segment, and the nodes are the HTML elements within the segment.
\end{itemize}

Using these models, the aesthetic score is computed by summing the size of the symmetric difference between each edge's labels in the \gls{sm} and \gls{ism} of the original page and transformed page~\cite{mfix}.

This is a multi-objective optimization problem, and thus we define the total fitness function as follows.
\begin{equation*}
F(\mathbb{X}) = \alpha \times E(U(\mathbb{X})-80) + \beta A(\mathbb{X})
\end{equation*},
where $\alpha$ and $\beta$ are negative and positive weight factors that could be tuned to obtain the best performance;
and $E(Y)$ is a function defined as:
\begin{equation*}
E(Y)=\left\{\begin{matrix}
G(Y) & \text{if~ } Y<0 \\ 
Y-1 & \text{otherwise} 
\end{matrix}\right.
\end{equation*}
We explain the high-level idea of the above fitness function as follows. First, it is worth noting that the maximum mobile-friendliness score given by Google Insight Tool is $100$ and that the usability score should be at least $80$ to guarantee the page's friendliness as indicated in~\cite{mfix}.
Motivated by this observation, we design the first item in the fitness function (i.e., the one representing the usability score) such that its value is significantly large when the usability score is smaller than the threshold, which is $80$. Therefore, $G(Y)$ should be a reciprocal function that increases rapidly compared to a linear function. The selection of $G(Y)$ is discussed in detail in Section \ref{sec:evaluation}

Note that as $\alpha$ is negative real number and $\beta$ is positive real number, maximizing $U(\mathbb{X})$ and minimizing $A(\mathbb{X})$ correspond to minimizing $F(\mathbb{X})$. Accordingly, the objective now is to minimize the fitness function $F(\mathbb{X})$.

\section{Proposed Algorithms}
\label{sec:proposal}
The core of our method is a \glsentrylong{pso} that repairs web pages by determining the optimal values of the fitness function as proposed in Section \ref{sec:formulation}. \gls{pso} iteratively tries to improve candidate solutions, here dubbed particles, by moving these particles around in the search space according to simple mathematical formulae over the particles' position and velocity. Using \gls{pso} instead of random searches helps to leverage knowledge about the changing trends of particles, improving and directing particles in the next iterations. \gls{pso} often provides high-quality solutions, but it may have a low-convergence rate which can be time-consuming. To further tackle this issue, we also propose a second approach, which optimizes for running time. This second approach is based on the Tabu search algorithm, which shows a higher convergence rate than \gls{pso}, as will be discussed in Section \ref{sec:evaluation}.

In the following, we describe the proposed algorithms in detail.
\subsection{PBRA: A  \gls{pso} Based Repair Algorithm}
\begin{algorithm}                      
\caption{PBRA}          
\label{algo:pbra}                           
\begin{algorithmic}[1]                    
\REQUIRE The set of segments and their corresponding issues and  \gls{pdg}s, $S_i, I_{i}^{j}, \operatorname{PDG}{\langle S_i,I_{i}^{j} \rangle} (i=\overline{1,n}; j\in{1, 2, 3})$ 
\ENSURE Repair values for the root nodes of  \gls{pdg}s
\STATE $x \leftarrow$ a candidate is suggested by  \gls{gmft} \cite{gmft}
\STATE Initializing the first generation consisting of $m$ particles $P_0=\left\{P_{1}^{0}, P_{2}^{0}, \ldots, P_{T}^{0}\right\}$ by using Gaussian distribution
\STATE $gBest \leftarrow F(P_{1}^{0})$
\FOR{$P^{0}_t \in P_0$}
    \STATE $v_0{(P_t)} \leftarrow$ random number in $(0,1)$
    \STATE $pBest_t \leftarrow P_{t}^{0}$
    \IF{$F(pBest_t) \leq F(gBest)$}
        \STATE $gBest \leftarrow pBest_u$
    \ENDIF
\ENDFOR
\STATE $k \leftarrow 0$
\WHILE{The termination condition has not been met}
    \FOR{$P_{t}^{k} \in P^k$}
        \STATE $v_{k+1}(P_{t}) \leftarrow \operatorname{Update}(v_{k}(P_{t}),pBest_t, gBest, P^{k}_t)$
        \STATE $P_{t}^{k+1} = P_{t}^{k} + v_{k+1}(P_{t})$
        \IF{$F(P_{t}^{k+1}) < F(pBest_t)$}
            \STATE $pBest_t \leftarrow P_{t}^{k+1}$
        \ENDIF
        \IF{$F(pBest_t) \leq F(gBest)$}
            \STATE $gBest \leftarrow pBest_t$
        \ENDIF
    \ENDFOR
    $k \leftarrow k+1$
\ENDWHILE
\RETURN $gBest$
\end{algorithmic}
\end{algorithm}
In this section, we leverage  \glsentrylong{pso} approach to determine the optimal values for CSS properties that need to be adjusted. 
Basically,  \gls{pso} approach simulates the behaviours of bird flocking. 
In  \gls{pso}, every single solution can be seen as a \emph{bird}, and we call it as a \emph{particle}.
The goodness of every particle is measured by a fitness value based on a fitness function. 
All of the particles have velocities that direct the flying of the particles by following the current optimum particles. 
\gls{pso} starts with a set of random particles and searches for the optima by updating generations. 
In each iteration, each particle is updated by following the best values found so far. 
Algorithm~\ref{algo:pbra}, presents the details of our proposed algorithm, which we will describe more in detail below.

Let us assume that we have a set of $m$ particles denoted as $\mathbb{P} = \left \{P_1, ...., P_T\right \}$. The algorithm then follows the steps below.

\begin{itemize}
\item \textbf{Step 1:}
The first step in the process of a  \glsentrylong{pso} is the generation of an initial population. Each particle of the initial population represents a possible solution to the problem. The initialization process is often designed to ensure the diversity of the population, which plays an important role in population-based algorithms. In mobile-friendly problems, we create the first particle by using the suggested value of  \gls{gmft} \cite{gmft}. Then, we generate the initial population by perturbing adjustment factors based on a Gaussian distribution around the original values of the first particle.
\item \textbf{Step 2:} Let $gBest$ denote the best value obtained so far by all particles in the population, and let $pBest_t$ denote the best value that the $t$-th particle (i.e., $P_{t}$) achieved so far. If the current value of the $t$-th particle is better than $pBest_t$, then $pBest_t$ is replaced by the current value of the $t$-th particle. The algorithm then searches for all particles to find the best candidate with the highest score and assigns it to $gBest$. 

\item \textbf{Step 3:} The algorithm leverage information obtained from previous generations to improve the quality of particles by adjusting particles direction based on $gBest$ and $pBest_t$, which is determined in step 2.
In detail, the velocity and updating the value for every particle are defined as follow. The velocity of particle $P_{t}$ at iteration $k+1$ (denoted as $v_{k+1}(P_{t})$) is determined as
\begin{equation*}
\begin{split}
v_{k+1}(P_{t}) &= w \times v_{k}(P_{t}) + c_1 \times r_1 \times \left( pBest_t - P^{k}_t\right) \\
&+ c_2 \times r_2 \times \left( gBest - P^{k}_t\right)
\end{split}
\end{equation*}
where $v_{k}(P_{t})$ is the velocity of particle $P_{i}$ at the previous iteration.
Based on the velocity, particle $P_{t}$ updates its value as follows.
\begin{equation*}
P_{t}^{k+1} = P_{t}^{k} + v_{k+1}(P_{t})
\end{equation*}
\item \textbf{Step 4:}
Step 2 is repeated until the termination condition is satisfied. The final value of $gBest$ is then deemed as optimal solution. In this work, we used the number of evaluations, i.e., the number of calls to the fitness functions, as our termination condition.
\end{itemize}
\subsection{TERA: a Time Efficient Repair Algorithm}
In this section, we present the detail of a repair algorithm based on Tabu search. 
The insight of our algorithm is that we start with a repair candidate that is suggested by \gls{gmft} \cite{gmft}. Then, we attempt to iteratively improve repair candidates over generations, following a genetic programming approach \cite{gp}. Algorithm~\ref{algo:tera} shows more details of our proposed algorithm.
\begin{algorithm}                      
\caption{TERA}          
\label{algo:tera}                           
\begin{algorithmic}
\REQUIRE The set of segments and their corresponding issues and \gls{pdg}s, $S_i, I_{i}^{j}, \operatorname{PDG}{<S_i,I_{i}^{j}>} (i=\overline{1,n}; j\in{1, 2, 3})$ 
\ENSURE Repair values for the root nodes of \gls{pdg}s
\STATE $x \leftarrow$ a candidate is suggested by  \gls{gmft} \cite{gmft}
\STATE $sBest$ $\leftarrow x$
\STATE $bestCandidate$ $\leftarrow x$
\STATE $tabuList$ $\leftarrow$ [ ]
\STATE $tabuList$.push($x$)
\WHILE{!\textit{termination condition}}
    \STATE $sNeighborhood$ $\leftarrow$ getNeighbors($bestCandidate$)
    \FOR{$sCandidate$ $\in$ $sNeighborhood$}
        \IF{$\operatorname{F}(sCandidate) \leq \operatorname{F}(bestCandidate)$ \textbf{and} $sCandidate$ $\notin$ $Tabulist$}
            \STATE $bestCandidate$ $\leftarrow$ $sCandidate$
        \ENDIF
    \ENDFOR
    \IF{F($bestCandidate$) $\leq$ F($sBest$)}
        \STATE  $sBest$ $\leftarrow$ $bestCandidate$
    \ENDIF
    \STATE $tabuList$.push(\textit{bestCandidate})
    \IF{$tabuList$.size() $>$ $size_{tabu}$}
        \STATE $tabuList$.removeFirst()
    \ENDIF
\ENDWHILE
\RETURN $sBest$
\end{algorithmic}
\end{algorithm}
We explain Algorithm~\ref{algo:tera} step by step below.
\begin{itemize}
\item \textbf{Step 1:} First, the algorithm creates a repair candidate is suggested by \gls{gmft} \cite{gmft}, assigning this candidate as the best candidate denoted by \emph{sBest}. It then inserts \emph{sBest} into \emph{TabuList} which is a list consisting of all the candidates that has been visited. 
 \item \textbf{Step 2:} Let \emph{bestCandidate} be the best repair candidate found in the previous iteration, and suppose that \emph{bestCandidate} is represented by \[\text{\emph{bestCandidate}}=\{x_{1}^{1}, x_{1}^{2}, x_{1}^{3}, ...., x_{n}^{1}, x_{n}^{2}, x_{n}^{3}\},\] where $x_{i}^{j}$ is the value for repairing issue $I_j$ of segment $x_i$. The algorithm then determines neighbors of \emph{bestCandidate} whose values fall into the following range: $\{x_{1}^{1} \pm \delta, x_{1}^{2} \pm \delta, x_{1}^{3} \pm \delta, ...., x_{n}^{1} \pm \delta, x_{n}^{2} \pm \delta, x_{n}^{3} \pm \delta\}$. Among the neighbors, it then assigns the one with the highest fitness as \emph{bestCandidate} of the current iteration. If \emph{bestCandidate} achieves a better fitness than the current \emph{sBest} then \emph{sBest} is replaced by \emph{bestCandidate}. \emph{bestCandidate} is then inserted into the \emph{TabuList}.
\item \textbf{Step 3:} Step  2 is repeated until the termination condition is matched. In this work, we used the number of evaluations, i.e., the number of calls to the fitness functions, as our termination condition.
\end{itemize}
\subsection{Implementation Detail}
We implemented our approach in Java, consisting of approximately 5000 lines of code. For identifying the mobile-friendly problems in a web page, we used well-established tools such as \gls{gmft} \cite{gmft} and \gls{psit} \cite{PSIT} APIs. For evaluating aesthetics, we identify segments in a web page and build the \glsentrylong{sm} and \glsentrylong{ism} by building a DOM tree following a similar method in \cite{mfix} with Chrome browser v60.0 and Selenium WebDriver.

\section{Experimental Result}
\label{sec:evaluation}
\begin{table*}[!h]
\centering
\caption{Statistics of real-world web-pages}
\resizebox{2.0\columnwidth}{!}{
\begin{tabular}{c l l c l l}
\\
\hline
\textbf{ID} & \textbf{Url} & \textbf{Categories} & \textbf{ID} & \textbf{Url} & \textbf{Categories} \\ 
\hline 
1 & https://www.discogs.com & Arts & 20 & https://www.wowprogress.com & Games\\
2 & https://xkcd.com  & Arts  & 21 & https://bulbagarden.net & Kids and teens\\
3 & http://www.wiley.com & Shopping & 22 & http://lolcounter.com & Kids and teens\\
4 & http://forum.gsmhosting.com/vbb & Home & 23 & http://www.bom.gov.au & Kids and teens \\
5 & https://www.irs.gov & Home  & 24 & http://onlinelibrary.wiley.com & Business\\
6 & https://travel.state.gov & Home  & 25 & http://aamc.org & Health \\
7 & https://arxiv.org & Science  & 26 & https://www.fragrantica.com & Health \\
8 & https://bitcointalk.org & Science  & 27 & http://us.battle.net & Kids and teens \\
9 & http://coinmarketcap.com & Science  & 28 & http://blizzard.com & Kids and teens \\
10 & http://www.intellicast.com & Science  & 29 & http://drudgereport.com & News\\
11 & https://www.ncbi.nlm.nih.gov & Science  & 30 & https://www.irctc.co.in & Regional \\
12 & http://sigmaaldrich.com & Science  & 31 & http://dict.cc & Reference \\
13 & http://www.weather.gov & Science  & 32 & https://www.leo.org & Reference  \\
14 & https://boardgamegeek.com & Games & 33 & http://correios.com.br & Society \\
15 & http://www.finalfantasyxiv.com & Game & 34 & http://www.flashscore.com & Sports\\
16 & http://www.mmo-champion.com & Home & 35 & http://letour.fr & Sports  \\ 
17 & http://www.nexusmods.com & Games & 36 & http://rotoworld.com & Sports \\
18 & http://nvidia.com & Games  & 37 & http://us.soccerway.com & Sports \\
19 & http://www.square- enix.com & Games & 38 & http://myway.com & Computers  \\
\hline
\end{tabular}
}
\vspace{-5pt}
\label{tab:datasets}
\end{table*}

In this section, we empirically compare our approach with the well-known automated webpage repair algorithm MFix~\cite{mfix} on a dataset of real-world webpages. Below, we describe our experimental methodology, including the data set and evaluation metrics, followed by research questions and numerical results.

\subsection{Dataset and Evaluation metrics}
\label{subsec:dataEval}
 To evaluate all algorithms, we perform the experiments on the dataset Alexa used in MFix~\cite{mfix}. The dataset contains 38 real-world subject web pages collected from the top 50 most visited websites across all seventeen categories. Table \ref{tab:datasets} presents the dataset's details, such as the URL and category of each web page. The datasets can be publicly accessed in the Github repository of MFix \footnote{\url{https://github.com/USC-SQL/mfix/tree/master/ICSE_paper_data/subjects}}. In terms of evaluation, we use two evaluation metrics proposed by S.Mahajan et al. \cite{mfix}:
\begin{itemize}
    \item \textbf{Usability Score}: We use \glsentrylong{psit} \cite{PSIT}, which assigns pages a score in the range of 0 to 100, with 80 being an entirely mobile-friendly page, to evaluate the usability of a web page. The higher the score, the better the usability of a web page is.
    \item \textbf{Aesthetic Score}: The aesthetic score is a metric proposed in~\cite{mfix} for measuring how beautiful a web page looks. Based on the assumption that the original web pages have the aesthetic layout, it is accounted for by the relative visual positioning within the segments of a page, such as \glsentrylong{sm} and \glsentrylong{ism}, between machine-patched and original web pages. In this way, the lower the aesthetic score, the more the page is aesthetically pleasing. Note that, in this study, we chose to use automated validation, which measures the differences between the repaired page and the original page, instead of a manual human to account for aesthetics. Although we acknowledge the imperfection of automated evaluation, human evaluation is expensive, e.g., the cost to hire professional developers to ensure the quality of the manual evaluation, and still this evaluation method suffers from human biases. Automated validation, on the other hand, is less expensive and does not suffer from human biases. That is a reason why we choose automated validation instead of manual human validation.
\end{itemize}
For each web page, we run TERA, PBRA, and MFIX, each for ten times, and calculate the average to mitigate the non-determinism inherent in the approximation algorithms. To ensure fairness between algorithms, we also use the same number of evaluations (number of fitness calls) for all algorithms. We also perform Wilcoxon signed-rank statistical tests~\cite{Wilcoxon1992} to ensure that the performance difference between our proposed solutions and the baseline is statistically significant. Our experiments were conducted on a Macbook Pro machine with an Intel Core i7 2.2 GHz and 16 GB of RAM, running macOS Mojave 10.14.5

\begin{table}[]
    \caption{Default setting for PBRA}
    \centering
    \resizebox{1\columnwidth}{!}{
    \begin{tabular}{l l l}
        \hline
         \textbf{Parameters} & \textbf{Symbol} & \textbf{Value}  \\
         \hline
         Population size & $T$ & 10 \\
         Number of evaluations & $E$ & 150 \\
         Inertial coefficient (start)& $\omega_{s}$ & 0.9\\
         Inertial coefficient (start)& $\omega_{e}$ & 0.4\\
         Acceleration coefficient (persional)& $c_{1}$ & 0.5\\
         Acceleration coefficient (global)& $c_{2}$ & 0.5 \\
         Coefficient of randomness & $r_{1}$, $r_{2}$& $[0,1]$\\
         Neighbor range & $\Delta$ & 0.3 \\
         Fitness Function & $G(Y)$ & $-Y^{2}$ \\
         \hline
         
    \end{tabular}
    }
    \label{tab:pbra_default}
\end{table}

\begin{table}[]
    \centering
    \caption{Default setting for TERA}
    \resizebox{1\columnwidth}{!}{
    \begin{tabular}{l l l}
        \hline
         \textbf{Parameters} & \textbf{Symbol} & \textbf{Value}  \\
         \hline
         Neighbor size & $size_{neighborhood}$ & 10 \\
         Number of evaluations & $E$ & 150 \\
         Size of tabu list & $size_{tabu}$ & 5 \\
         Neighbor range & $\delta$ & 0.3 \\
         Fitness Function & $G(Y)$ & $-Y^{2}$ \\
         \hline
         
    \end{tabular}
    }
    \label{tab:tera_default}
\end{table}

\begin{figure*}[bt]
    \centerline{\includegraphics[width=1\linewidth]{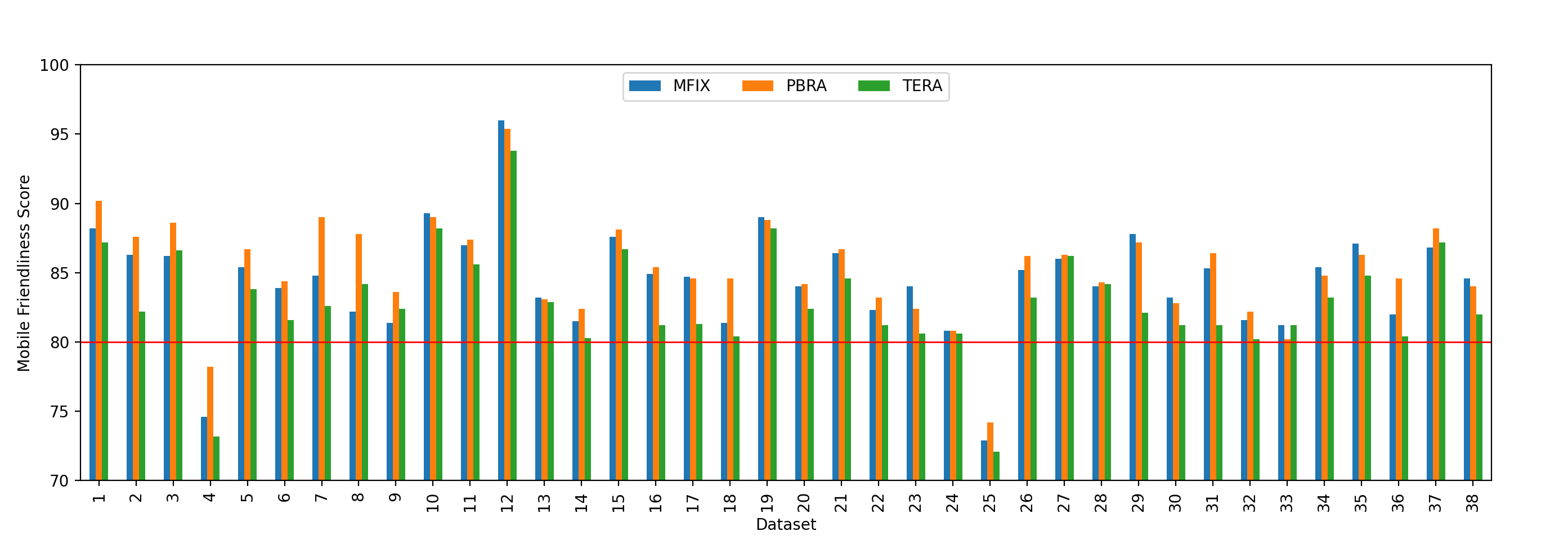}}
    \caption{\textbf{Comparison of TERA, PBRA and MFix in Usability.} The  higher  the  usablity  score,  the  more  the mobile friendly. Values above the red horizontal line drawn at 80 indicate that the GMFT considers a page to be mobile friendly.}
    \label{fig:comparison_usa}
    \vspace{-5pt}
 \end{figure*}
 
\begin{figure*}[bt]
    \centerline{\includegraphics[width=1\linewidth]{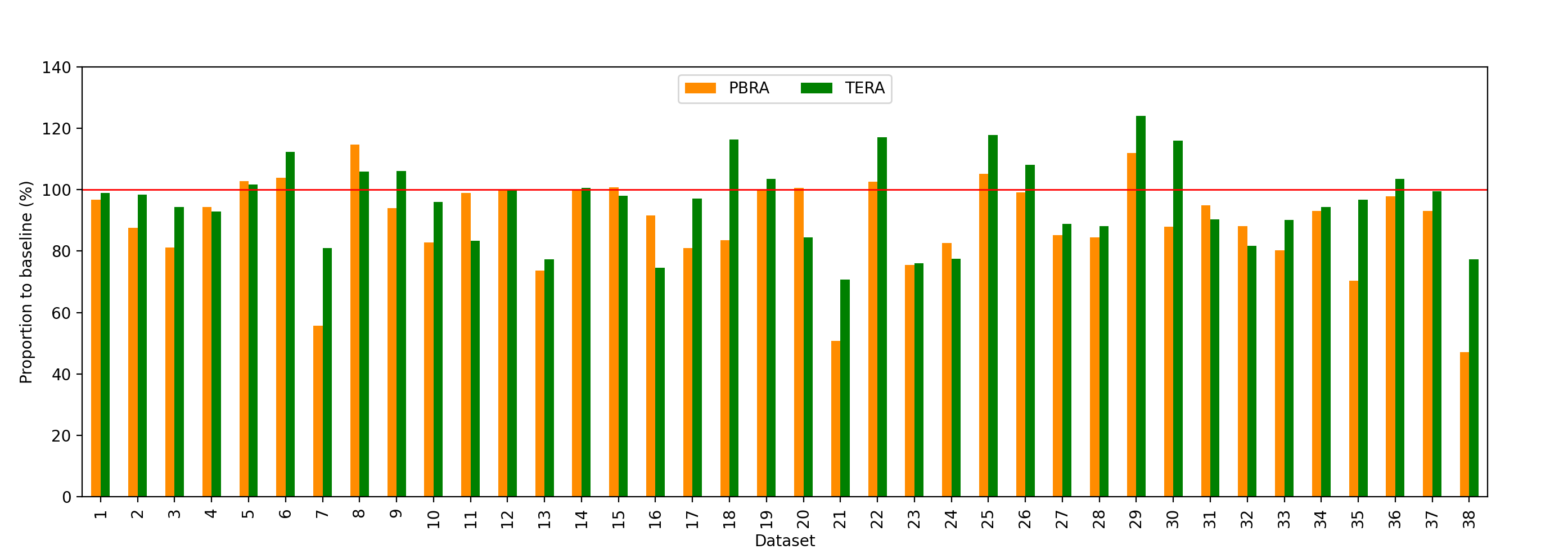}}
    \caption{\textbf{Comparison of TERA and PBRA versus MFix in aesthetics.} The  lower  the  aesthetic  score,  the  more  the page is aesthetically pleasing. Values below the red horizontal line drawn at 100\% indicate improvements over baseline.}
    \label{fig:comparison_aes}
    \vspace{-5pt}
\end{figure*}

\subsection{Research Question}
We answer four research questions as described below.

\vspace{0.2cm}
\noindent \textbf{RQ1: How effective is our approach in repairing mobile-friendly problems in web pages as compared to MFix?}  In this research question, we investigate the effectiveness of our approach, as compared to MFix, to generate repairs in the benchmark dataset that we describe in Section \ref{subsec:dataEval}. Note that our comparison relies on mobile-friendly tasks as well as aesthetic layout tasks.

\vspace{0.2cm}

\noindent \textbf{RQ2: How do different fitness functions impact on PBRA and TERA effectiveness?}
By default, our approach uses a fitness function with function $G(Y)$ is quadratic. In this research question, we compare different fitness functions to explain why we selected the quadratic function as default.

\vspace{0.2cm}

\noindent \textbf{RQ3: How efficient is TERA as compared to PBRA?} In this question, we investigate the efficiency between TERA and PRBA by performing a comparison on their convergence rate. 
\vspace{0.2cm}

\noindent \textbf{RQ4: How do different population sizes impact on PBRA effectiveness?}
By default, our approach selects an initial population size of 10 for our approach. In this research question, we compare different values of initial population size to explain why 10 is the best option. 

\subsection{Numerical result}
\begin{figure*}
\centering
\begin{subfigure}[H]{0.49\textwidth}
\includegraphics[width=\textwidth]{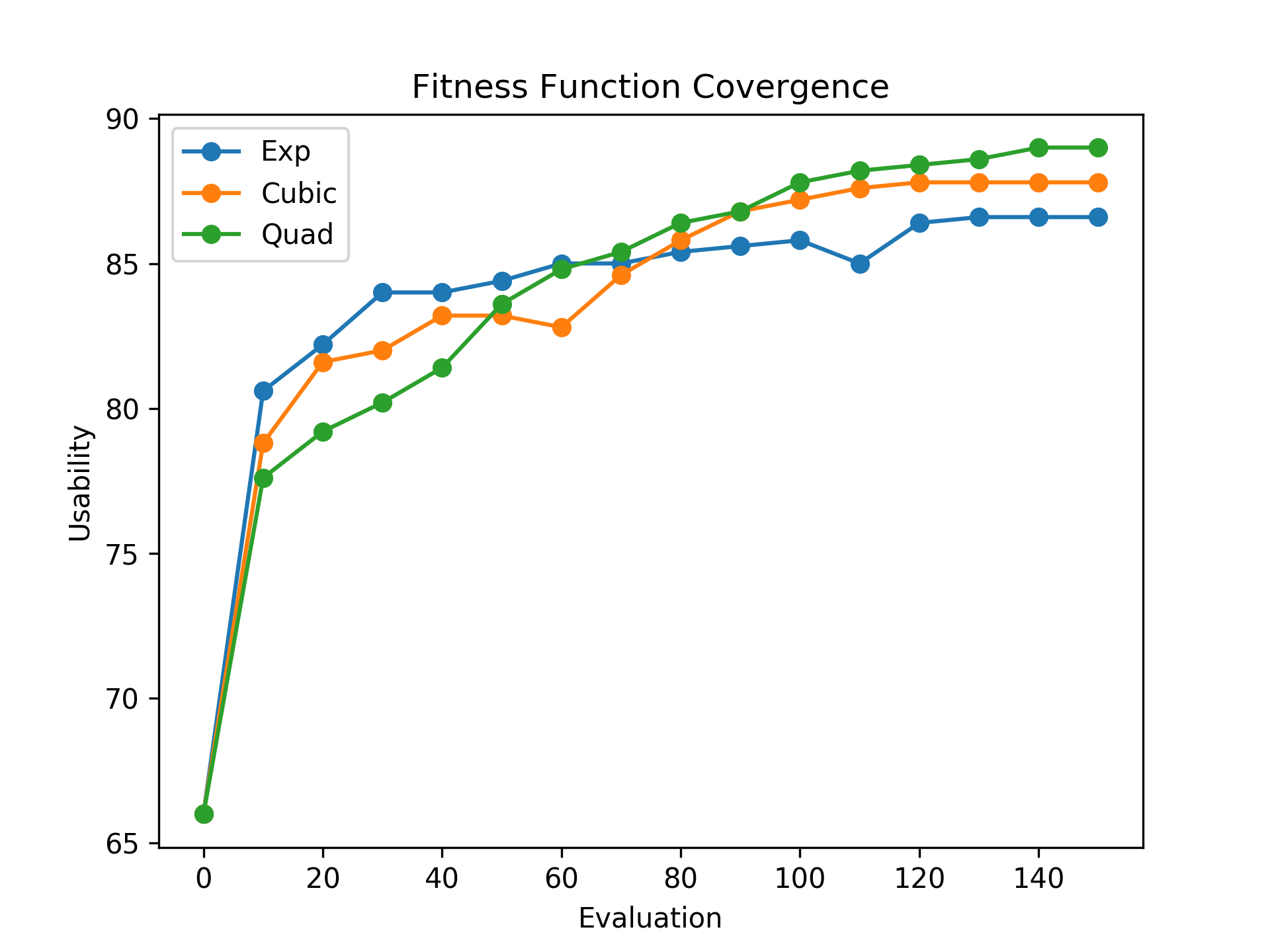}
\caption{Usability}
\end{subfigure}
\begin{subfigure}[H]{0.49\textwidth}
\includegraphics[width=\textwidth]{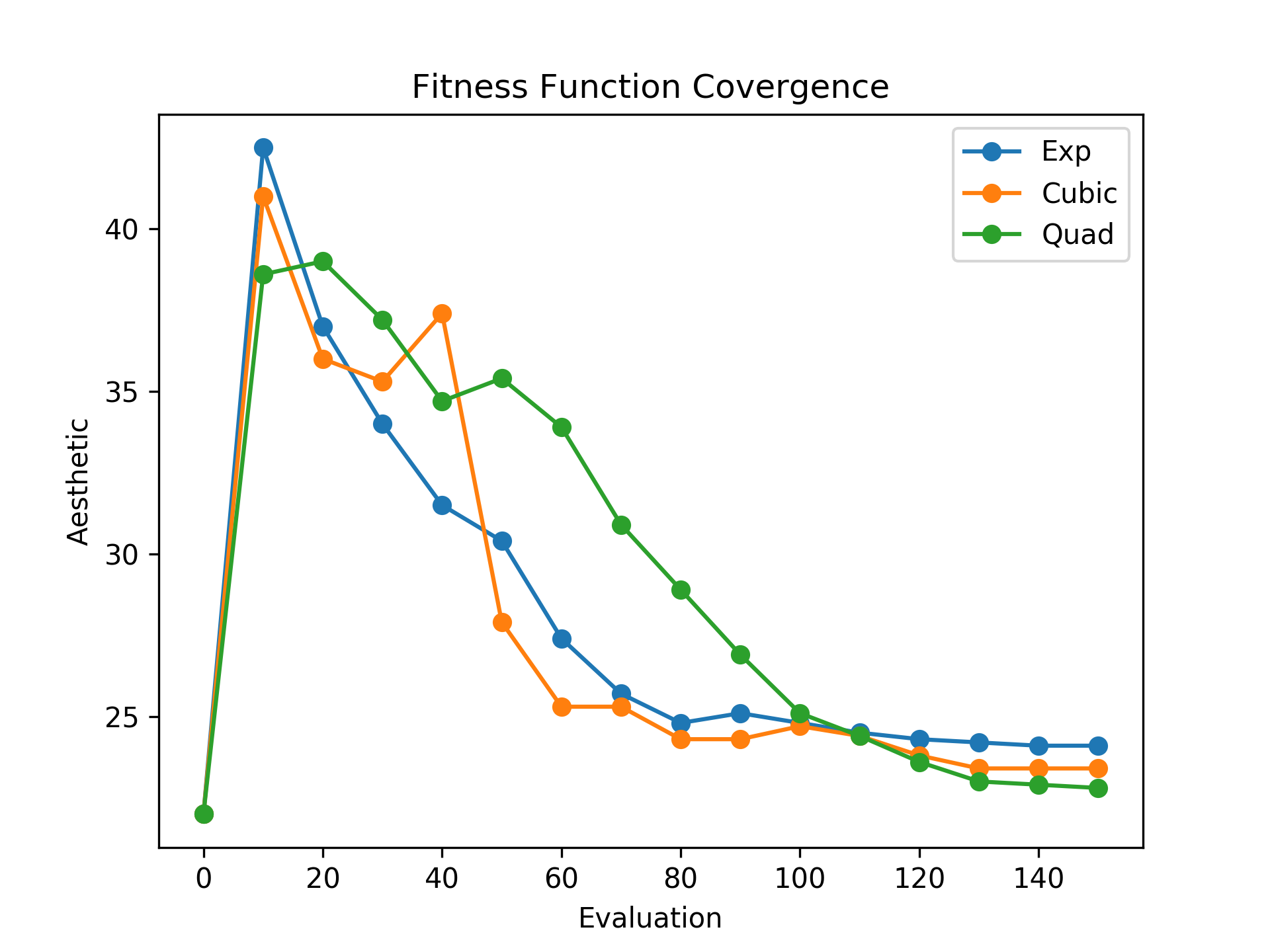}
\caption{Aesthetics}
\end{subfigure}
\caption{Impact of fitness function on convergence rate.\label{fig:fitness}}
\end{figure*} 
\par \textbf{RQ1: Our approach's effectiveness as compared to MFix.} To answer RQ1, we compare the effectiveness of our approach including PBRA and TERA (with default settings on Table \ref{tab:pbra_default} and Table \ref{tab:tera_default}) with a well-known automated repair technique in mobile-friendly problems, MFix~\cite{mfix}. We also performed statistical tests to study the significance of the improvements by our approaches over the baseline.
Rationale behind default settings on table \ref{tab:pbra_default} and table \ref{tab:tera_default} will be explained by RQ3 and RQ4.

\par Experiment results show that PBRA outperforms MFix by generating repairs that have much better usability and aesthetic scores. Note that, as we described in Section~\ref{subsec:dataEval}, the higher the usability score, the better the usability of a web page, and the lower the aesthetic score, the more aesthetically pleasing a web page is. From Figure \ref{fig:comparison_usa}, we can clearly find that our approach, namely PBRA, outperforms MFix in 26 out of 38 pages and achieves almost the same performance with MFix in other subject web pages. The results show that PBRA is significantly better than TERA and MFix in the mobile-friendliness task. 

\par Figure \ref{fig:comparison_usa} represents the usability of the repaired web pages by using TERA, PBRA, and MFix. The red line shows a threshold value of 80 - which is the minimum \gls{psit} score to be considered for good usability by \gls{gmft}~\cite{gmft}. We can see that most of the repairs pass the \gls{psit} threshold (36 out of 38 subjects).

\par \par The numerical results concerning aesthetic score are depicted in Figure \ref{fig:comparison_aes}. In Figure \ref{fig:comparison_aes}, we show the relative performance of our approaches versus the baseline technique MFix by dividing the aesthetics score of PBRA-patched and TERA-patched pages by that of the MFix-patched pages. We do this to normalize the results because of a wide range of aesthetics' value, e.g., the aesthetics' value can range from anywhere greater than zero. Figure \ref{fig:comparison_aes} shows that PBRA improves the aesthetic score significantly, outperforming MFix on 29 out of 38 subjects in aesthetics. TERA also shows considerably better results than MFix on 24 out of 38 subjects.

\par We also perform Wilcoxon signed-rank test \cite{Wilcoxon1992} to investigate the performance difference between our approaches and MFix. The results point out that PBRA is significantly better than both TERA and MFix in usability scores (with p-values of 1.4 e-07 and 0.003, effect-size of 0.627 and 0.222, respectively). PBRA and TERA also are better than MFix in aesthetics (with p-values of 0.004 and 0.041, effect-size of 0.991 and 0.435, respectively).


\begin{figure*}
\centering
\begin{subfigure}[b]{0.49\textwidth}
\includegraphics[width=\textwidth]{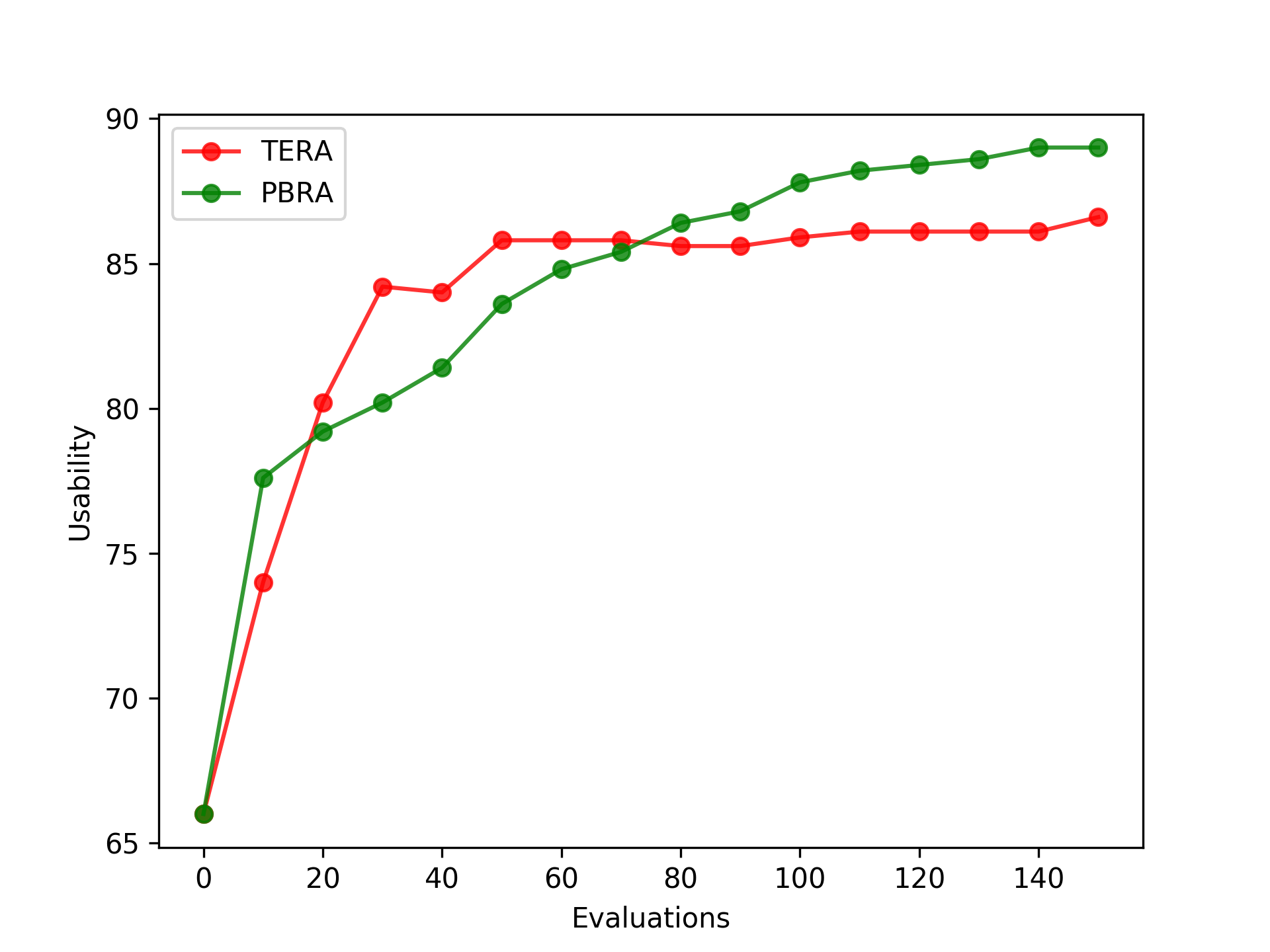}
\caption{Usability}
\end{subfigure}
\begin{subfigure}[b]{0.49\textwidth}
\includegraphics[width=\textwidth]{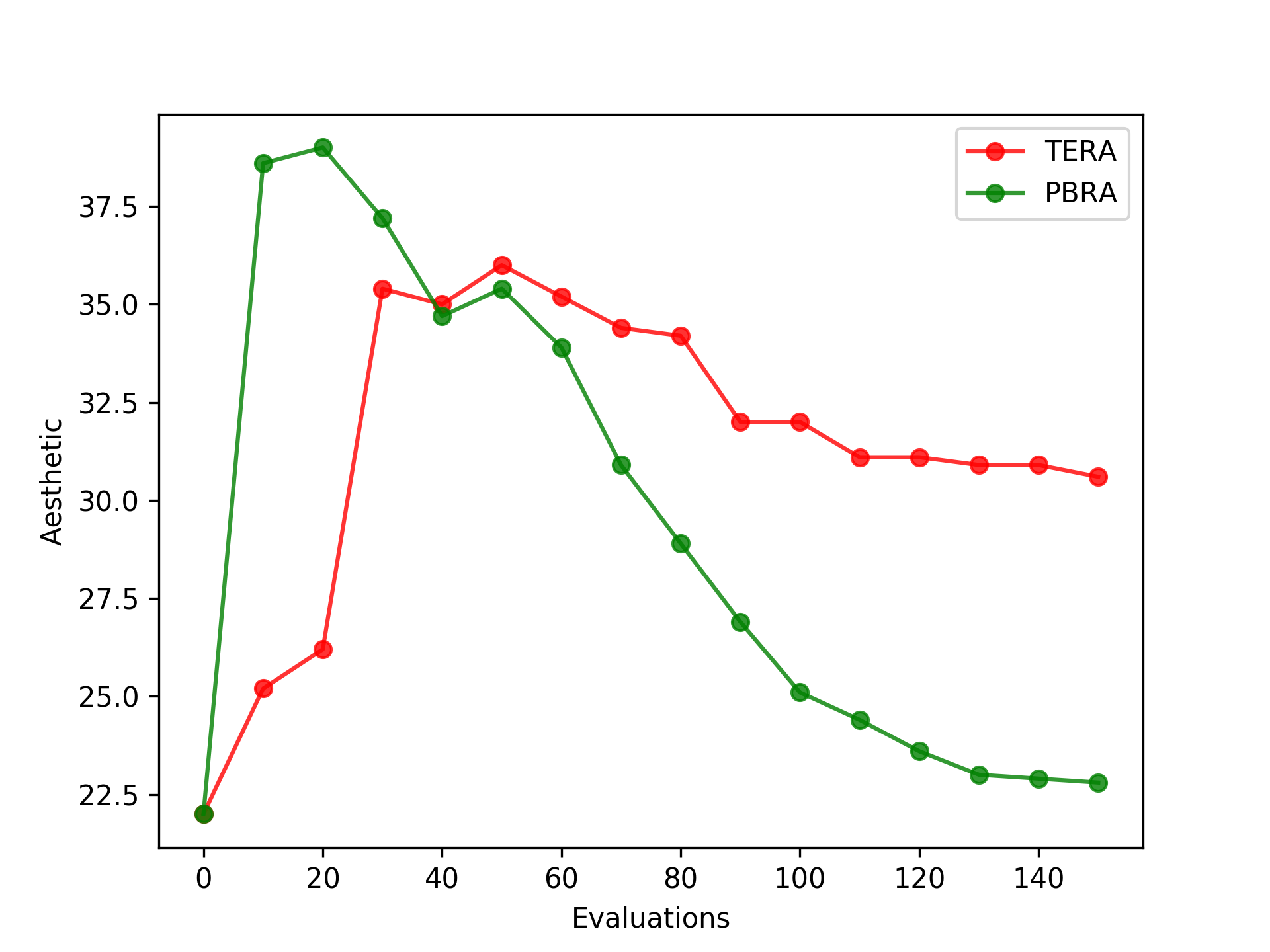}
\caption{Aesthetics}
\end{subfigure}
\caption{Comparison on covergence rate of TERA and PBRA \label{fig:convergence}}
\end{figure*}

\vspace{0.2cm}

\begin{figure*}[h]
\centering
\begin{subfigure}[b]{0.49\textwidth}
\includegraphics[width=\textwidth]{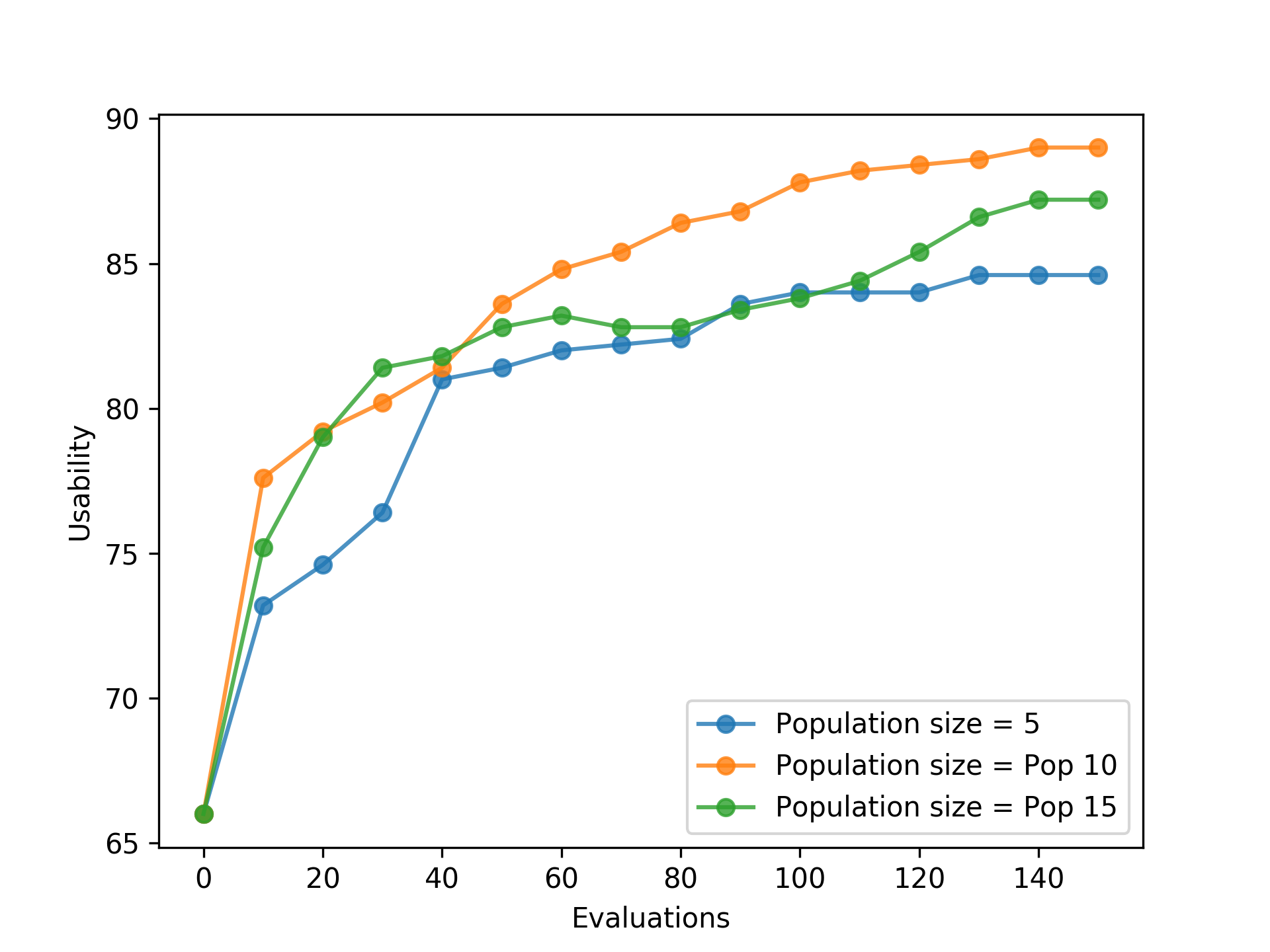}
\caption{Usability}
\end{subfigure}
\begin{subfigure}[b]{0.49\textwidth}
\includegraphics[width=\textwidth]{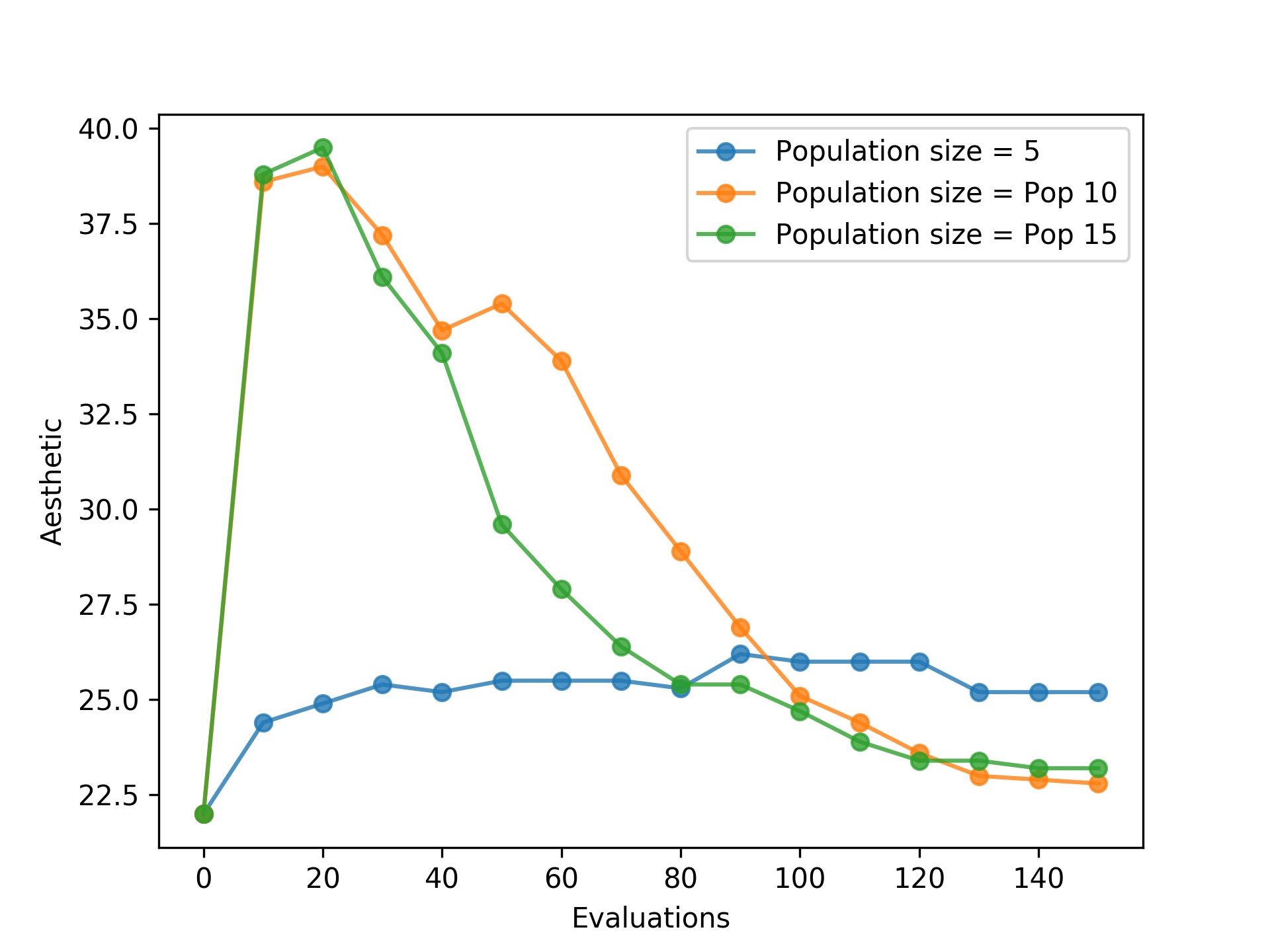}
\caption{Aesthetic}
\end{subfigure}
\caption{Impact of population size on covergence rate\label{fig:Convergence_PopSize}}
\end{figure*}

\par \textbf{RQ2: Impacts of the fitness function.}
In Section \ref{sec:formulation}, we introduce a novel fitness function to combine usability score and aesthetic score. In the formula, $G(Y)$ function plays an important role in making adjustments to speed up the search for friendly patches. However, the $G(Y)$ function also needs to be designed to avoid ignoring potentially good candidate patches.

We have chosen some functions which satisfy the conditions of G(Y) (in Section \ref{sec:formulation}) to experiment with, such as:
\begin{itemize}
    \item Exponential function $G(Y) = e^{-Y}$
    \item Quadradtic function $G(Y) = -Y^{2}$
    \item Cubic function $G(Y) = Y^{3}$
\end{itemize}

We compare the effectiveness of the fitness functions on two tasks: convergence rate, i.e., how fast a fitness function helps the approaches to converge, and overall result. In the overall effect, from Figure \ref{fig:fitness}, we can clearly find that the quadratic function is the most effective, followed by the cubic function and exponential function.
In convergence rate tasks, Figure \ref{fig:fitness} shows that our approach with exponential fitness function converges quickly with values exceeding 80. 
 
Indeed, exponential fitness function can converge well just only within a small number of evaluations of candidate solutions. However, when the number of evaluations is increased, there is no significant improvement. Our experiments show that cubic and quadratic fitness functions need more computation time to achieve better results. One potential reason could be that the exponential fitness function focuses on improving usability score to exceed the threshold so that it ignores potentially good candidate patches. In conclusion, the quadratic function has the best fitness function for both PBRA and TERA. We, therefore, have chosen the quadratic function as the default fitness function for our approach. 

\begin{figure*}
\centering
\begin{subfigure}[b]{0.49\textwidth}
\includegraphics[width=\textwidth]{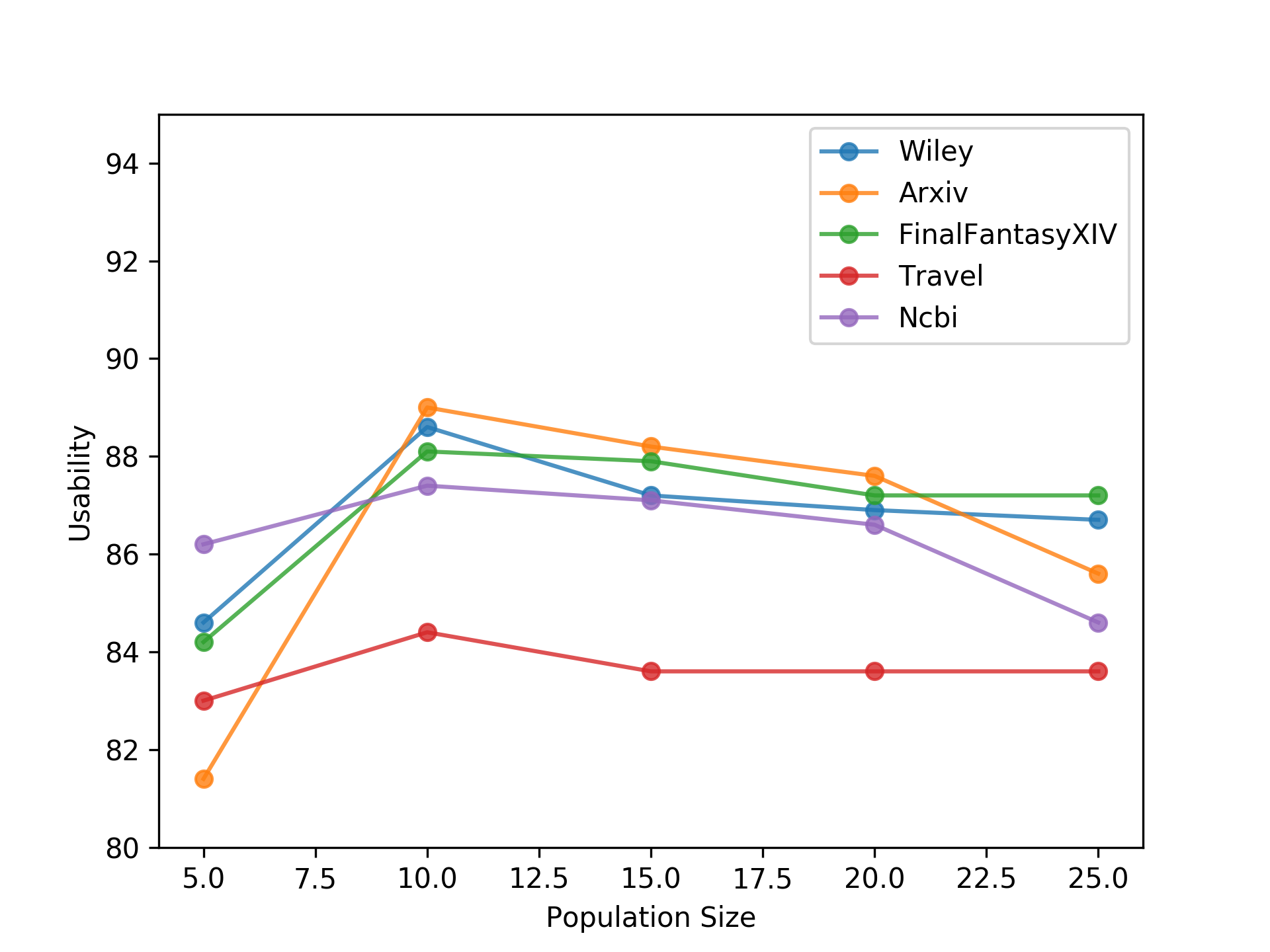}
\caption{Usability}
\end{subfigure}
\begin{subfigure}[b]{0.49\textwidth}
\includegraphics[width=\textwidth]{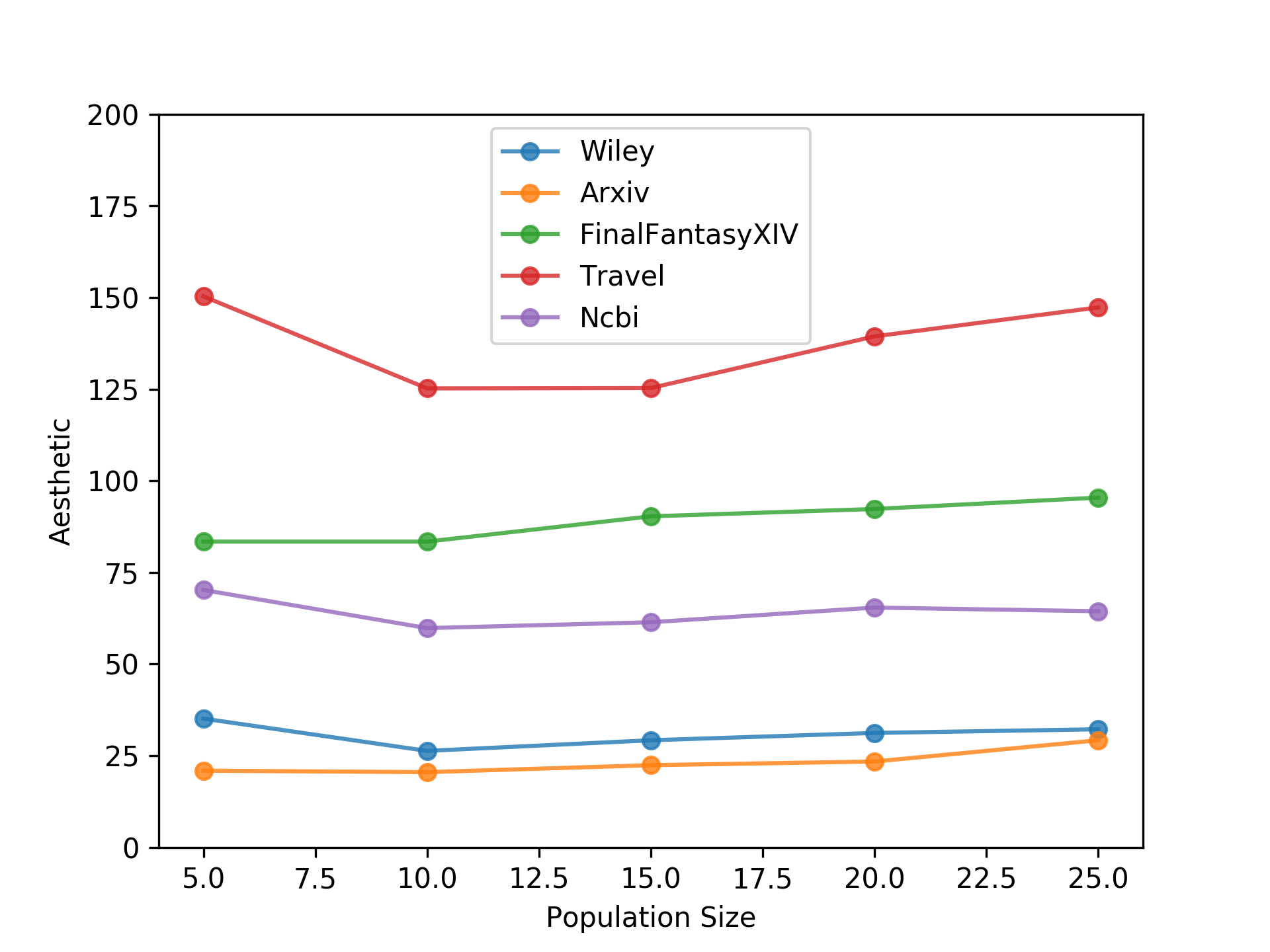}
\caption{Aesthetic}
\end{subfigure}
\caption{Impact of population size on overall result\label{fig:PBRA_PopSize}}
\end{figure*}

\par \textbf{RQ3: Convergence rates of PBRA and TERA.}
The first RQ has demonstrated that PBRA outperforms TERA (36 out of 38 on the mobile-friendliness test and 30 out of 38 on aesthetics). In this RQ, we further investigate the convergence rates of both PBRA and TERA to assess the efficiency of the algorithms.

Our experiments on all pages suggest a similar trend as shown in Figure~\ref{fig:convergence}, which is the result for the page http://www.wiley.com.
Figure~\ref{fig:convergence} depicts the values of usability score, aesthetic score, and fitness over generations.
As shown, TERA converges more quickly than PBRA, but PBRA achieves better values than TERA.
\par Concerning usability score, TERA converges when the number of evaluations is $50$, while PBRA needs $150$ evaluations to converge (see Figure~\ref{fig:convergence}a).
TERA only needs $30$ evaluations to reach score 80 - which is the threshold to pass \gls{gmft} \cite{gmft}, while PBRA needs $50$ evaluations. This means that TERA and PBRA can guarantee a high level of friendliness for repaired pages, but TERA consumes less running time. 
\par The convergence times of TERA and PBRA concerning the aesthetic score are similar, as depicted in Figure \ref{fig:convergence}b.
Both of them converge when the number of evaluations is $150$.
The converged value of PBRA is much smaller than that of TERA. Specifically, PBRA attains the aesthetic score that equals to $86\%$ that of TERA. This result means that PBRA can provide a better aesthetic than TERA.

\vspace{0.2cm}
\par \textbf{RQ4: Impacts of Population Size} Population size plays an essential role in population-based meta-heuristic methods. A small population size may not be sufficient to find good results, while a large population size may hamper the convergence rate. Therefore, we perform experiments with different population sizes on two tasks: the convergence rate and overall result. Figure \ref{fig:Convergence_PopSize} shows the impact of the population size on the convergence rate of the algorithm. From the figure, we can clearly find that the population size of 10 is the most suitable value of the approach, while the population size of 5 and population size of 15 shows the worse result. 
Small populations are more likely to experience the loss of diversity over time because the mating between individuals with similar genetic structures occurred regularly. Therefore, the search space of the algorithm is narrowed, which usually leads to poor results like the result of the population size of 5 in this case. In contrast, large populations could make the algorithm consume more computation time in finding an optimal solution. That is the reason why algorithms with a population size of 15 have poor results. This observation is further reinforced through the experiment on the overall result, which is shown in Figure \ref{fig:PBRA_PopSize}, with five subjects and five different values of population size.

\section{Threats to validity} \label{sec:threats}
\textbf{External Validity.} Threats to external validity correspond to the generalizability of our findings. Our evaluation dataset contains 38 real-world web pages. Still, these subjects may not represent all possible cases in reality, and thus our results may not generalize. We tried to mitigate this by selecting top-ranked websites from various categories. We plan to experiment on a larger dataset in the future. 

\vspace{0.2cm}
\textbf{Internal Validity.} Threats to internal validity refers to errors in our implementation and experiments. To mitigate this risk, we have carefully examined our code and experiments to avoid potential errors. We have also repeated our experiments several times to ensure that the results reported are correct.

\vspace{0.2cm}
\textbf{Construct Validity.} 
Threats to construct validity correspond to the suitability of our evaluation metrics. Following MFix, we consider web pages as mobile-friendly if it passes a threshold value of 80, which is the minimum \gls{psit} score for good usability. These ratings are, however, subject to the \gls{psit} evaluation criteria only. Although \gls{psit} is a popular tool to assess usability, we plan to also include other tools to more thoroughly assess usability.

\section{Conclusion} \label{sec:conclusion}
In this paper, we introduced automated approaches based on a meta-heuristic algorithm for repairing mobile-friendly problems in web pages. Our approaches strive for improving mobile-friendliness (usability) while minimizing layout disruption (aesthetic) at the same time. To achieve this, we designed a new novel fitness function to allow our search algorithms to find better repairs more efficiently and effectively. Our evaluations on a data set of 38 real-world defective webpages show that our approaches generate good mobile-friendly patches for 94\% of the subjects and significantly outperform existing state-of-the-art baseline technique MFix. We have also studied the impact of different fitness functions and parameters of the meta-heuristic algorithm on the overall result and convergence rate. Overall, experimental results are promising and indicate the usefulness of our approaches, which may support developers in designing mobile-friendly web pages. 

In the future, we plan to improve the effectiveness and efficiency of our solution further by designing better fitness function to better optimize for both usability and aesthetic. We also plan to curate more data to expand our experiments on a larger dataset containing many more real-world projects. Furthermore, we plan to integrate our approach into development pipeline to interactively help developers develop better web pages.

\section{Acknowledgements}
We would like to thank Mahajan et. al for sharing their implementation of MFix~\cite{mfix} and guiding us through the experimental setup for MFix.



\bibliographystyle{IEEEtranS}
\bibliography{references.bib}

\end{document}